\documentclass[sigconf]{acmart}

\acmConference[ISSTA 2022]{ACM SIGSOFT International Symposium on Software Testing and Analysis}{18-22 July, 2022}{Daejeon, South Korea}
\AtBeginDocument{%
  \providecommand\BibTeX{{%
    \normalfont B\kern-0.5em{\scshape i\kern-0.25em b}\kern-0.8em\TeX}}}

\usepackage{colortbl}
\usepackage{bbding}
\usepackage{subfigure}
\usepackage{multirow} 
\setcopyright{acmcopyright}
\copyrightyear{2018}
\acmYear{2018}
\acmDOI{XXXXXXX.XXXXXXX}

\begin{document}

\title{Catch Me If You Can: Blackbox Adversarial Attacks on Automatic Speech Recognition using Frequency Masking}

\author{Xiaoliang Wu}
\affiliation{%
  \institution{University of Edinburgh}
  \city{Edinburgh}
  \country{UK}}
\email{X.Wu-53@ed.ac.uk}

\author{Ajitha Rajan}
\affiliation{%
  \institution{University of Edinburgh}
  \city{Edinburgh}
  \country{UK}
}
\email{arajan@ed.ac.uk}

\renewcommand{\shortauthors}{Wu and Rajan}


\begin{abstract}
Automatic speech recognition (ASR) models are prevalent, particularly in  applications for voice navigation and voice control of domestic appliances. The computational core of ASRs are deep neural networks (DNNs) that have been shown to be susceptible to adversarial perturbations; easily misused by attackers to generate malicious outputs. 
To help test the security and robustnesss of ASRS, we propose techniques that generate blackbox (agnostic to the DNN), untargeted adversarial attacks that are portable across ASRs. This is in contrast to existing work that focuses on whitebox targeted attacks that are time consuming and lack portability. 

Our techniques generate adversarial attacks that have no human audible difference by manipulating the audio signal using a psychoacoustic model that maintains the audio perturbations below the thresholds of human perception. We evaluate portability and effectiveness of our techniques using three popular ASRs and two input audio datasets using the metrics - \texttt{Word Error Rate (WER)} of output transcription, \texttt{Similarity} to original audio, attack  \texttt{Success Rate} on different ASRs and \texttt{Detection score} by a defense system. We found our adversarial attacks were portable across ASRs, not easily detected by a state-of-the-art defense system, and had significant difference in output transcriptions while sounding similar to original audio. 
\end{abstract}




\maketitle
\section{Introduction}
\label{sec:intro}
Automatic speech recognition models (ASRs) are widely used in a variety of applications, such as mobile virtual assistants (Siri, Google Assistant), in-vehicle voice navigation and voice smart home appliances like Alexa and Google Home with built-in voice assistants. Figure~\ref{fig:asr_structure} shows the structure of a typical ASR that takes as input an audio signal and transcribes the speech in the audio to text. Owing to the prevalence of ASRs in our daily lives, their security and integrity pose a great concern.   
The computational core of ASRs are deep neural networks (DNNs) that have been shown to be susceptible to adversarial perturbations; easily misused by attackers to generate malicious outputs~\cite{2017burger,nichols2017tvexamples,yuancommandersong}.

\paragraph{Existing work on ASR adversarial attacks.}
Adversarial perturbations\footnote{Also referred to as Adversarial examples or Adversarial attacks.} were first presented by Szegedy et al. to demonstrate the lack of robustness in DNN models -- a small
perturbation of an input may lead to a significant perturbation of the output of a DNN model~\cite{szegedy2013intriguing}. This vulnerability can be exploited by adversaries to augment the original input with a crafted perturbation, invisible to a human but sufficient for the DNN model to misclassify this input. 
This influential work triggered several research contributions in the computer vision domain that generate  adversarial attacks for testing security and robustness of vision tasks~\cite{goodfellow2014explaining, kurakin2016adversarial, moosavi2017universal}. Research on the use of adversarial attacks on ASRs is, however,  only just emerging, and 
can be classified along two dimensions, \\ 
\textbf{1. Un-targeted or Targeted} The aim of un-targeted adversarial audio is to make an ASR model incorrectly transcribe speech while sounding similar to original input, while the aim of targeted adversarial attack is to cause an ASR model to output a specific transcription (target) injected by an adversary. This paper focuses on un-targeted adversarial attack. \\
\textbf{2. Whitebox or Blackbox Threat Model} In a whitebox threat model, the adversary assumes knowledge of the internal structure of the ASR model, while in a blackbox threat model, the adversary can only probe the ASR with input audio and analyze the resulting transcription.  We use a blackbox threat model. 

Most existing methods~\cite{carlini2016,carlini2018,qin2019,Yakurarobust} for ASR adversarial attack generation are \emph{targeted and whitebox}.  These methods suffer from one or more of the following drawbacks (1) Whitebox assumption is not practical and lacks portability since commercial ASR application developers do not typically reveal the internal workings of their systems, 
(2) time taken to generate attacks is considerable and cannot be used in real-time. 
, and (3) poor quality audio in attacks makes them easily detectable by defense techniques like ~\cite{carlini2018,universal}. Existing few methods~\cite{didyouhear,taori2019targeted} for \emph{blackbox, targeted} attacks suffer from the drawback of intractable number of queries to the ASR, that are time-consuming and impractical.  
\emph{Blackbox untargeted} attacks that do not require knowledge of the internal NN structure or query access for text output would address the above limitations and the only known technique was proposed by Abdullah et al. in 2020~\cite{abdullah2019hear}. To create adversarial audio, they decompose the original audio and remove components with low-amplitude that they believe will not affect audio comprehension. Although interesting, their approach does not strive to ensure the adversarial and original audio sound similar and difference in transcribed texts is not measured. Additionally, the ability of their attacks to bypass defense systems is not effective.  


\paragraph{Proposed Attack Generation} We propose a blackbox un-targeted attack generation approach that is faster, more portable across ASRs, and robust to a state-of-the-art defense than Abdullah et al. 
Our approach for attacking ASRs 
uses a psychoacoustics concept called frequency masking that determines how sounds interfere and mask each other. We manipulate masked (or inaudible) components of the original audio in such a way that their spectral density is different but they remain masked. Such a manipulation ensures the adversarial attack is indistinguishable from the original but has the potential to change the resulting transcription.  
We propose three attack generation approaches centered around this idea -- \texttt{Griffin Lim Reconstruction (GL),Original Phase (OP)} and \texttt{Deletion (DE)}. Additionally, to help increase similarity to the original audio, we provide the option of selectively introducing perturbations to a small fraction of audio frames rather than all of them. Our approach provides three frame selection options -- \texttt{Random, Important} and \texttt{All}. Among them, the \texttt{Important} option identifies the frames that cause the most change to output text when set to zero and we then introduce perturbations to just these important frames. 

We evaluate our approach on three different ASRs -- Deepspeech \cite{deepspeech2014}, Sphinx \cite{Sphinx} and Google cloud speech-to-text API, using two different input audio datasets -- Librispeech \cite{librispeech} and Commonvoice \cite{commonvoice:2020}. We assess the effectiveness of our approaches for attack generation and frame selection using the metrics - \texttt{WER, Similarity}, attack \texttt{Success Rate} and \texttt{Detection score}. We also compare our approach with a targeted whitebox state-of-the-art (SOTA) method~\cite{carlini2018} and an untargeted blackbox SOTA method~\cite{abdullah2019hear}. It is worth noting that the scale of our evaluation is much bigger than existing work~\cite{carlini2018,qin2019,abdullah2019hear} as we use different audio datasets and ASRs. We find our approach that uses \texttt{OP} or \texttt{DE} for attack generation combined with \texttt{Important} or \texttt{All} frame selection was effective at attacking all three ASRs. Our techniques were $312\times$ faster than the whitebox targeted SOTA, 
and $7\times$ faster than blackbox targeted SOTA method. 
The defense system, Waveguard~\cite{waveguard}, was less effective at detecting attacks generated with our techniques compared with the other two SOTA methods.


\noindent In summary, the contributions in this paper are as follows:
\begin{enumerate}
    \item A novel approach for untargeted blackbox adversarial attack generation on ASRs based on frequency masking.
    \item Frame selection option to selectively perturb frames in an audio.
    \item Extensive empirical evaluation of the attack generation and frame selection options within our approach on three ASRs and two audio datasets. We also compare performance against SOTA whitebox and blackbox techniques.
\end{enumerate}
\noindent The source code for our approach can be found at: \\ \indent \href{https://anonymous.4open.science/r/lalalala-9DEE}{https://anonymous.4open.science/r/lalalala-9DEE}.

\section{Background}
We present a brief description of a typical ASR model and the frequency masking concept used in our approach.
\label{sec:background}
\subsection{Automatic Speech Recognition (ASR)}
\label{sec:ASR}
\begin{figure*}[htbp]
\includegraphics[scale=0.665]{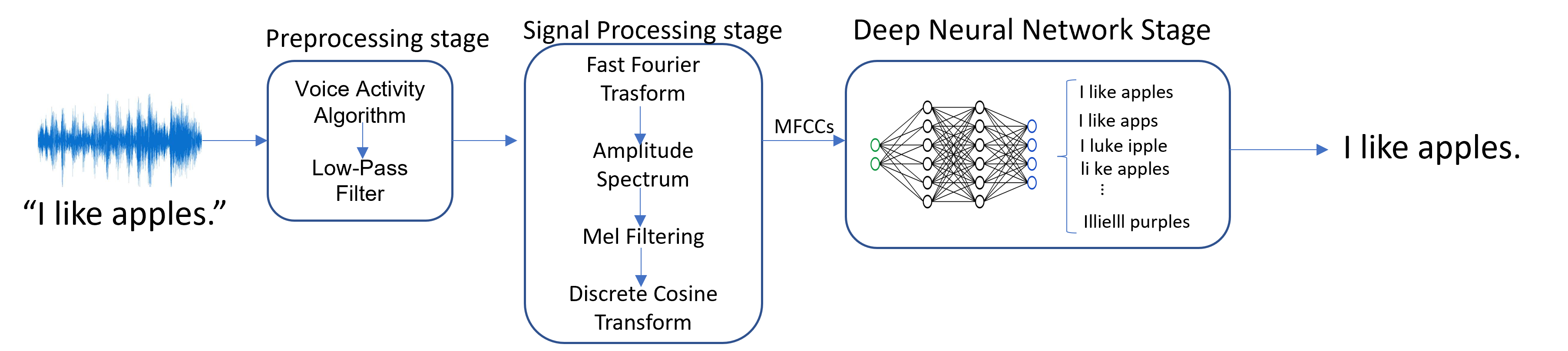}
\caption{Pipeline showing Stages in a typical Automatic Speech Recognition (ASR) System.}
\label{fig:asr_structure}
\end{figure*}

Structure and workflow within a typical ASR is shown in Figure~\ref{fig:asr_structure}. Most current ASRs comprise the following stages when transcribing an input audio to a text output. 

\subsubsection{Preprocessing}
This step removes high-frequency noise in the audio.
A voice activity algorithm is used to detect human voice parts in a given input audio and then passes it through a low-pass filter to remove high-frequency noise that is inaudible to humans.

\subsubsection{Signal Processing stage}
\label{sec:signal}
Output from this stage is audio features that are subsequently used by a deep neural network. 
In the signal processing stage, the audio signal in the time domain is sampled into frames with a certain sampling rate(like 16000HZ and 8000HZ) and every frame is converted to the frequency domain using Fast Fourier Transform. The result of this step is a complex matrix, where the real part of the matrix is the amplitude information of the frame, and the imaginary part is the phase information. The phase spectrum is discarded, and only the amplitude spectrum is retained. This amplitude spectrum is the expression of the audio in the frequency domain, which details different frequencies and corresponding intensities in the frame.
Subsequent steps in the ASR are completed on the basis of the amplitude spectrum. 

To extract audio features, the amplitude spectrum is passed through Mel filters and Discrete Cosine Transform (DCT). The output is Mel Frequency Cepstral Coefficient (MFCC), which is commonly used in ASRs as features of audios. Detailed description of this step can be found in~\cite{SOK}.

\subsubsection{Neural network prediction and output selection stage}
The extracted features from the audio are fed into a deep neural network (DNN), such as a Recurrent Neural Network, that then predicts a probability distribution of characters for every time step or audio frame. From the character sequence distributions, an output selection algorithm, such as Beam search, is used to select the most likely translated text as shown in Figure~\ref{fig:asr_structure}. More details on this stage can be found in~\cite{SOK}.

It is worth noting that much of the existing work on adversarial attacks against ASRs are aimed at the DNN stage (prediction stage) and typically use gradient-based optimization to minimize the difference between the target and output text~\cite{carlini2018,qin2019,Cal_masking}.  
In contrast, our approach for generating adversarial attacks does not rely on a target output text or query outputs from the ASR. We, instead, make changes to the original audio signal based on frequency masking of its components that is described in the next Section. 

\subsection{Frequency Masking and Masking Threshold Computation}
\label{sec:masking}
\begin{figure}[htbp]
\includegraphics[width=0.5\textwidth, height=150pt]{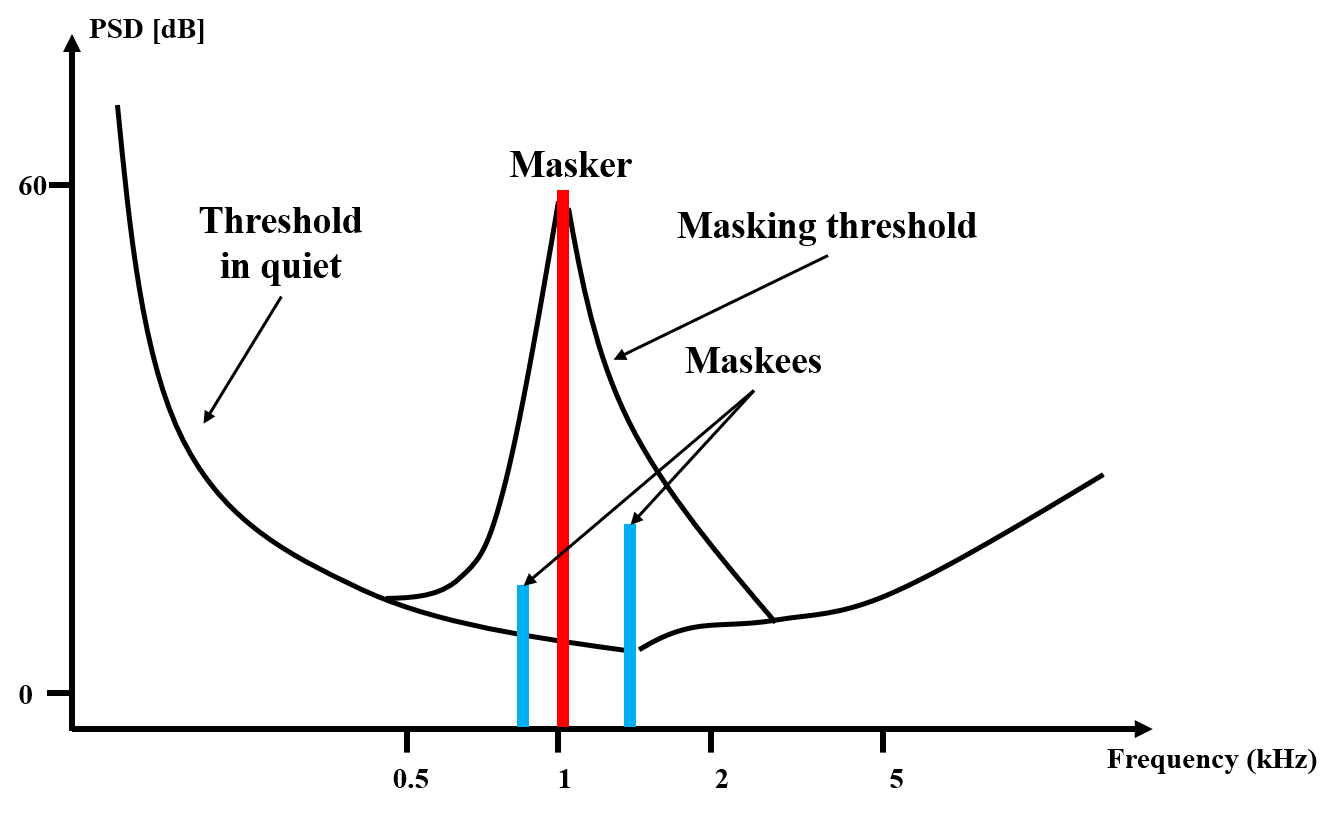}
\caption{Frequency masking phenomenon: the masker creates a \emph{masking threshold} in the nearby frequency domain such that other sounds below this threshold cannot be heard.}
\label{fig: masking_threshold}
\end{figure}
Frequency masking is a psychoacoustic phenomenon that occurs when the perception of a sound is affected and masked by the presence of another sound, distracting the ear from being able to clearly perceive the simultaneous sounds~\cite{Cal_masking}. For example, on a quiet night, consider that the sound  of chirping crickets is audible but in the presence of the TV sound, we stop hearing the crickets chirping as the TV sound masks it. In Figure~\ref{fig: masking_threshold}, the TV sound would be the \emph{masker} (seen as a red bar) that creates a masking threshold~\cite{Cal_masking} which is the minimum level at which other sounds in the same frequency frame can be heard. The chirping sound of the crickets falls below the masking threshold (seen as a blue bar) and therefore is not audible in the presence of the TV. The chirping sound in Figure~\ref{fig: masking_threshold} would be the \emph{maskee}. 

\paragraph{Masking Threshold Computation}
To calculate the masking threshold for a given audio, we need to first convert the audio from the expression in the time domain to the frequency domain (using FFT in Section~\ref{sec:signal}), then discard the phase information in the spectrum. We then use the amplitude information of the spectrum to calculate the log-magnitude power spectral density (PSD) of this audio. The PSD characterizes the energy distribution on a unit frequency, and is used widely to describe the frequency domain results of the signal~\cite{masking_threshold,Cal_masking}. The red and blue bars in Figure~\ref{fig: masking_threshold} represent the PSD (in dB) of maskers and maskees, respectively, for the given frequency bin.  According to~\cite{Cal_masking,qin2019}, maskers are  identified from the audio PSD using two conditions: the PSD of a masker should be greater than the absolute threshold of hearing (ATH), and  it must be the highest PSD estimate within a certain surrounding frequency range. After identifying the maskers, their respective masking thresholds will be computed using a two-slope function, described  in~\cite{masking_threshold}. If there are several maskers and associated masking thresholds, they will be combined into a global masking threshold for the audio like in~\cite{qin2019}. Once the maskers are identified, the other PSDs in the audio are labelled maskees. A more detailed description of the computation of masker, maskee and masking threshold can be found in~\cite{masking_threshold,qin2019}.

We use this masking phenomenon observed with simultaneous sounds to create adversarial audio that sounds similar to the original audio but has the potential to produce a different transcription. 
We achieve this by first taking the original audio that is composed of many sounds, identifying the maskers and maskees in it using the approach from~\cite{masking_threshold,qin2019} (red and blue bars in Figure~\ref{fig: masking_threshold}). We then manipulate the PSD of the maskees so it stays below the masking threshold, ensuring they are not audible, like in the original audio. Nevertheless, this manipulation can still affect the transcribed text. We create the adversarial audio by composing together the unchanged maskers and manipulated maskees.  
In terms of our earlier example with the TV sound and crickets chirping, we identify the TV sound as the masker and the chirping crickets as the maskee. We then manipulate the PSD of the cricket sound, staying within the masking threshold, to produce an adversarial audio that composes the TV sound with the manipulated chirping sound.  Section~\ref{sec:methodology} describes our approach and the techniques used for manipulation in detail. 

\subsection{Griffin-Lim Algorithm}
\label{sec:GLbackground}
To construct an adversarial audio from the maskers and manipulated maskees in the amplitude spectrum, we use 
the Griffin-Lim (GL) algorithm that helps reconstruct audio waveforms with a known amplitude spectrum but an unknown phase spectrum\cite{GL1984}. 
Steps in the algorithm are as follows:
(1) Randomly initialize a phase spectrum,
(2) Use this phase spectrum and the known amplitude spectrum to synthesize a new waveform through Inverse Short-Time Fourier Transform
(3) Use the synthesized speech to get new amplitude spectrum and new phase spectrum through Short-time Fourier Transform,
(4) Discard the new amplitude spectrum, 
(5) Repeat steps 2, 3, 4 for a fixed number of iterations.
Output is a waveform with an estimated phase spectrum and the known input amplitude spectrum. 

\section{Methodology}
\label{sec:methodology}
\begin{figure}[htbp]
\includegraphics[width=0.75\textwidth]{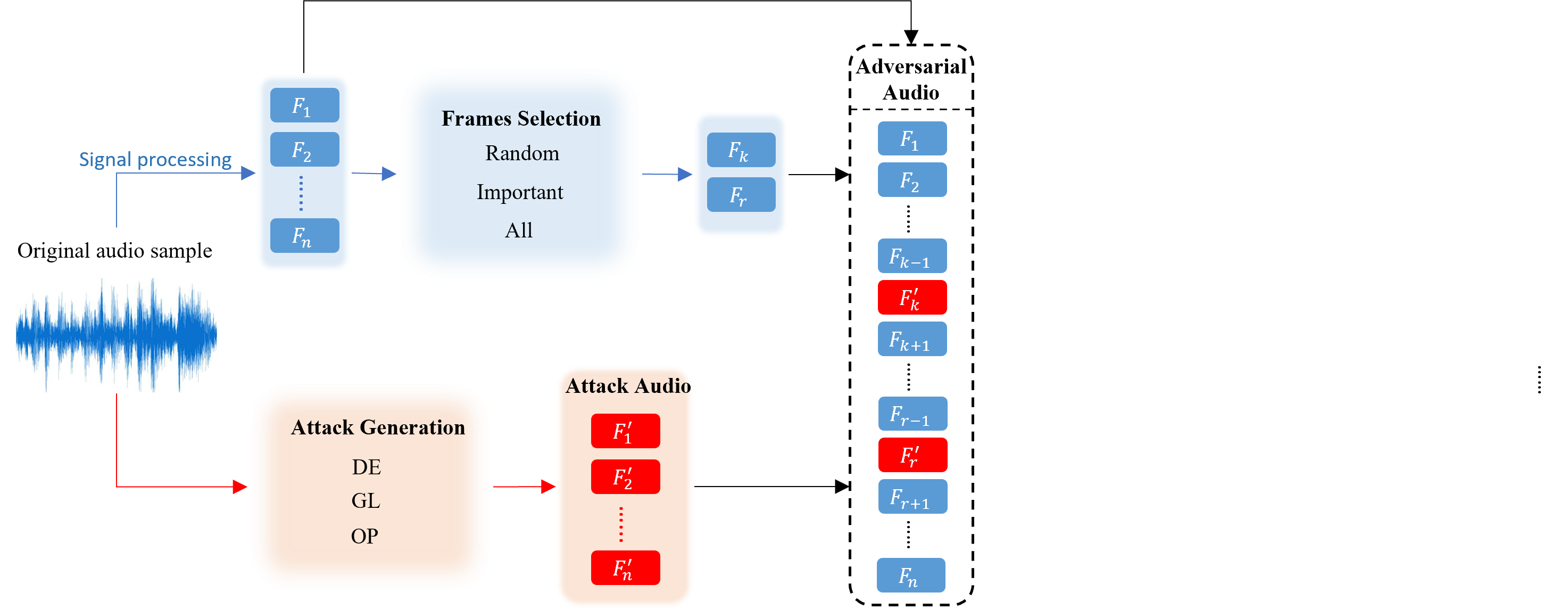}
\caption{Our approach for generating adversarial attacks comprises of three stages, 1. Frame Selection, 2. Attack generation and finally 3. Adversarial audio formed by combining information in the first two stages.}
\label{fig: framework}
\end{figure}
In this section, we propose techniques for generating adversarial attacks for ASRs. As seen in Figure~\ref{fig: framework}, our methodology has two important stages, 1. Audio Frame Selection and 2. Attack Generation.  The general workflow in our approach is as follows: Given an input audio example, we first select frames within it using one of the three techniques for audio frame selection -- \texttt{Random, Important} and \texttt{All}. Independently, we generate manipulated audio from the input audio using one of three attack techniques -- \texttt{GL Reconstruction (GL), Original Phase (OP), Deletion (DE)}. We then replace the selected frames in the original audio with corresponding manipulated audio frames while keeping the rest of the audio unchanged. The combination of original and manipulated audio frames forms the adversarial attack audio. 

\paragraph{Threat Model and Assumptions} The attack techniques in our approach assume a black-box threat model, in which an adversary has no knowledge of the internal workings or architecture of the target ASR model. We treat the ASR as a black-box to which we make requests in the form of input audio and receive responses in the form of transcriptions in text format. 
We also assume that an adversary can only make a limited number of requests to the target ASR. We also accommodate the scenario when the adversary cannot make any requests to the target ASR.  
Finally, we assume an over the line attack. This means that digital files are sent directly to the target ASR system for transcription, as opposed to playing back audio files over the air through speakers.

\subsection{Stage 1: Frame Selection}
As mentioned in~\ref{sec:signal}, the audio signal input to an ASR is sampled into frames in the signal processing stage. We explore generation of adversarial audio by modifying a subset of frames in the entire audio. We provide three approaches to select audio frames that will be later manipulated  -- \texttt{Random, Important} and \texttt{All}. We will start by describing the technique to select \texttt{Important} frames. 
\subsubsection{Important:} The rationale for selecting important frames is to restrict manipulation to a small number of significant frames. This allows the adversarial audio to remain similar to the original while still affecting the output transcription text. We define importance of frames based on the proportion of \texttt{WER} produced by masking that frame in the original audio. The steps involved in selecting important frames are as follows, 
\begin{enumerate}
    \item For every input audio example, record output translated text from ASR. 
    \item Pick one of the input audio examples. For every frame in the processed audio example, set it to zero (masked) while keeping the remaining frames unchanged. Record translated text using the ASR for the masked audio. 
    \item Compute \texttt{WER} between the masked and original output. Repeat this for all frames.  The frames that result in a non-zero \texttt{WER} are identified as important frames for that audio example. Magnitude of \texttt{WER} change for frame selection can be altered to suit needs.  
    \item Repeat Steps 2 and 3 for the remaining input audio examples.

\end{enumerate}
At the end of this process, every input audio example is associated with a list of important frames. 

\subsubsection{Random:}
To enable us to compare the effectiveness of only using important frames in frame selection, we also provide a means to select frames randomly. The number of frames selected for a given audio example is set to be the same as the number of important frames in that audio. 

\subsubsection{All:}
We simply use \emph{all} the frames from the manipulated audio generated in Stage 2 (see Section~\ref{sec:stage2}). Using \texttt{All} frames helps us assess how much \texttt{WER} was achievable. In addition it helps quantify the tradeoff in \texttt{WER} and \texttt{Similarity} when compared to frame selection with \texttt{Important} and \texttt{Random}. 

\begin{figure}[htbp]
\includegraphics[width=0.5\textwidth, height=180pt]{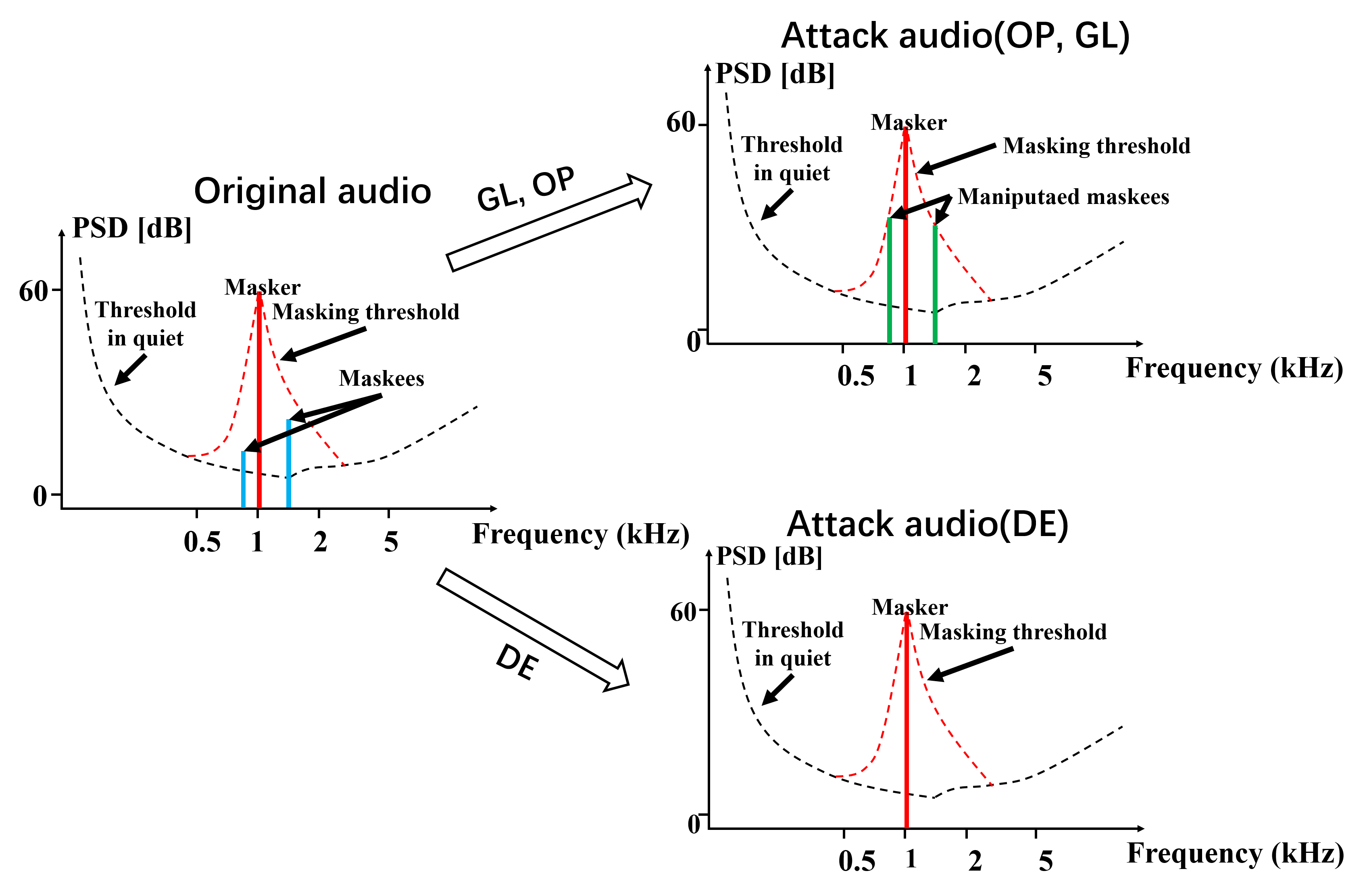}
\caption{Attack generation methods, GL and OP, increase the PSD of maskees to the masking threshold. Attack generation with DE suppresses the PSD of maskees to zero.}
\label{fig: method_3}
\end{figure}

\subsection{Stage 2: Attack Generation}
\label{sec:stage2}
We discuss three attack generation techniques -- GL, OP and DE,  that manipulate the amplitude spectrum of the input audio example using the concept of frequency masking, described in Section~\ref{sec:masking}.
We illustrate the manipulations in Figure~\ref{fig: method_3} and describe them in the Sections below. All three techniques take the input audio, generate audio frames in the frequency domain (obtained with sampling and fast fourier transform), with each frame having amplitude and phase information. For each frame, we compute the masking threshold, maskers and maskees using established techniques discussed in Section~\ref{sec:masking}

\subsubsection{GL Reconstruction (GL)}
\label{sec:GL}
As seen in the top part of Figure~\ref{fig: method_3}, \texttt{GL} (and \texttt{OP}) increases the PSD of all maskees (blue bars in the original audio) to the global masking threshold. Masker PSDs remain unchanged.  We then compute an updated amplitude based on the maskers and altered maskees PSD inversely~\cite{masking_threshold}\footnote{$Amplitude(k)=N\sqrt{10^{\frac{PSD(k))}{10}}}$, where $k$ is the index of the frequency bin and $N$ represents the length of frame. }. \texttt{GL} discards phase information of the input audio waveform. Instead, it estimates phase information using the GL reconstruction technique discussed in Section~\ref{sec:GLbackground}. The estimated phase information is combined with the updated amplitude information and is used to synthesize the attack audio through inverse FFT.   

\subsubsection{Original Phase (OP)}
The primary difference between the \texttt{OP} and \texttt{GL} technique is in the phase information. Estimating phase using the GL algorithm introduces distortion and lack of consistency across multiple runs. To avoid this problem, the \texttt{OP} technique retains phase information from the original audio. We believe using phase information from the original audio to synthesize the attack audio will make it more similar to the original audio. 

\subsubsection{Deletion (DE)}
Previous methods, \texttt{OP} and \texttt{GL}, ensure the attack audio sounds no different from the original input by increasing the PSD of the maskees up to the maximum limit (which is the masking threshold) for them to remain masked. The \texttt{DE} technique, on the other hand, suppresses the PSD of the maskees to the minimum value of zero which is akin to deleting them. This manipulation will not affect the audio perception as the masking threshold is unaffected. The \texttt{DE} technique, thus, deletes all maskee PSDs that are hidden under the masking threshold. Subsequently, we use the modified amplitude after deletion and combine it with the \emph{original phase} information from the input audio (similar to \texttt{OP}'s use of phase). We use inverse FFT as before to synthesize attack audio from the amplitude and phase information. 

\subsection{Stage 3: Combining Original and Attack Audio}
In this final stage, we create an adversarial attack by taking the original audio, replacing the selected frames (identified in Stage 1) with corresponding frames from the attack audio (generated in Stage 2). Other frames from the original audio are left unchanged. This modified version of the original serves as an adversarial attack. 

The source code for our adversarial attack generation approach, with the three attack generation and three frame selection methods, can be found at ~\href{https://anonymous.4open.science/r/lalalala-9DEE}{https://anonymous.4open.science/r/lalalala-9DEE}.

\section{Experiments}
We evaluate the effectiveness of our techniques, described in Section~\ref{sec:methodology}, using two different datasets -- (1) 1000 audio samples from Librispeech~\cite{librispeech} and (2) 200 audio samples from Commonvoice~\cite{commonvoice:2020}. We use three ASRs in our evaluation, namely,  Deepspeech~\cite{deepspeech2014}, Sphinx~\cite{Sphinx}, and Google ASR.  Our choice of datasets and ASRs were inspired by their use in related work for adversarial ASR attack generation~\cite{abdullah2019hear}\cite{carlini2018}\cite{qin2019}\cite{Zhang2017DolphinAttack}. We discuss the defense system used to assess the effectiveness of the adversarial attacks, evaluation metrics and the research questions in our experiments in the rest of this Section. 

\subsection{Detection and defense}
\label{sec:defense}
The ability to evade defense systems is an important measure of effectiveness for adversarial attacks. Defense systems have evolved to detect and defend a significant fraction of adversarial attacks.  
In our experiments, we  use a SOTA adversarial audio detection and defense system,  Waveguard~\cite{waveguard}, proposed by Hussain et al. in 2021. 
We chose Waveguard as our defense system as it is demonstrated to be faster, more effective and capable of detecting both targeted and untargeted attacks compared to existing detection techniques, like Temporal Dependency Detection Method~\cite{Temporal}. We report how well Waveguard performed (as an AUC score) in detecting adversarial attacks in our experiments.  

Attack detection within Waveguard is divided into two steps. The first step is to transform the input audio using one of several functions that are meant to preserve (or closely preserve) the transcription text. For example, a transformation may start by down-sampling the input audio, followed by up-sampling to the original sampling rate using interpolation. 
The second step is to compare the Character Error Rate(CER) between the transcription text for the original and transformed audio. If the difference between the texts is greater than a predefined threshold, then the input audio is classified as adversarial, and benign otherwise. 

\subsection{Evaluation Metrics}
\label{sec:metrics}
We use four metrics to measure the effectiveness of our techniques  -- \texttt{Word Error Rate (WER), Similarity, Success Rate} and \texttt{Detection score}. We are interested in generating adversarial attacks that sound similar to the the original audio (high \texttt{Similarity}) but produce a transcription different from the original (high \texttt{WER}).
Additionally, we would like the technique to be portable, i.e  generate adversarial attacks that are usable across several ASRs (high \texttt{Success Rate}). Finally, we want the generated attacks to be robust to get past SOTA defense systems, like Waveguard~\cite{waveguard} (lower \texttt{Detection score}).
We provide definitions of each of these metrics below. 

\paragraph{\texttt{WER}} is a common metric to evaluate the difference in ASR transcription from original versus adversarial audio~\cite{WER1}~\cite{WER2}.
\texttt{WER} is computed using Equation~\eqref{WER},
\begin{equation}
    \text{WER}=\frac{\text {Insertions}+\text {Substitutions}+\text {Deletions}}{\text {Total Words in Correct Transcript}}
\label{WER}
\end{equation}
\paragraph{Similarity} We use the widely used \texttt{PESQ} metric~\cite{PESQ} that measures quality of audio relative to a reference audio to assess similarity of adversarial audio to the original. The PESQ algorithm accepts a noisy signal, which in our case is the adversarial attack,  and an original reference signal, which is the input audio for our method. 
The PESQ score ranges from -0.5 to 4.5. The higher the score, the better the voice quality. According to~\cite{PESQover3}, audio quality is deemed ``good" when its \texttt{PESQ} score is above 3.0. We use this standard for classifying the quality of the adversarial audio. 
In this paper, we use \texttt{Similarity} metric to mean the \texttt{PESQ} score. 
\paragraph{\texttt{Success Rate}} shown in Equation~\eqref{success}, refers to the ratio of adversarial attacks that can successfully attack a given ASR. A successful attack, as defined by Abdullah et al~\cite{abdullah2019hear}, happens when the adversarial attack results in a non-zero WER with respect to the original transcription. 
\begin{equation}
    \text{Success Rate}=\frac{\text {Number of successful attacks}}{\text {Total number of adversarial attacks}}
    \label{success}
\end{equation}

\paragraph{\texttt{Detection score}} refers to the effectiveness of the Waveguard defense system in correctly classifying adversarial attacks. We use the area under the curve (AUC) metric, reported by Waveguard~\cite{waveguard},  to evaluate correct classification of adversarial attacks. 
The AUC score ranges from 0.0 to 1.0. 
We aim for a lower Waveguard AUC score or \texttt{Detection score} with our techniques. 

\subsection{Research Questions}
\label{sec:RQ}
We aim to answer the following research questions (RQs) in our experiments, \\
\noindent\textbf{RQ1:} Which frame selection method among \texttt{Random, Important, All} performs best? \\
We compare the \texttt{WER} and \texttt{Similarity} achieved by the different frame selection techniques across three different ASRs and two input audio datasets. Answering this research question will help us assess the value of selecting a subset of frames versus just changing the whole audio. 

\noindent\textbf{RQ2:} Which attack generation technique among \texttt{GL, OP, DE} performs best?\\
We compare the \texttt{WER}, \texttt{Similarity} achieved by the different attack generation techniques across three different ASRs and two different input datasets. We also measure \texttt{Time} taken by each technique. 

\noindent\textbf{RQ3:} Are the adversarial attacks portable across ASRs?\\
One of the primary selling points of our techniques is that they are blackbox and untargeted, and therefore agnostic to the structure and workings within ASRs. We validate this by evaluating the  \texttt{Success Rate} of the generated adversarial attacks across three different ASRs.

\noindent\textbf{RQ4:} Does our technique perform better than SOTA techniques?  \\
We selected representative and high-performing SOTAs in our comparison, namely a whitebox targeted technique proposed by Carlini et al~\cite{carlini2018}, and a blackbox technique by Abdullah et al~\cite{abdullah2019hear}. 

Carlini et al. generate adversarial attacks using Deepspeech ASR and the Commonvoice input dataset. To allow comparison, we use the same ASR and input dataset with our techniques. Owing to the targeted nature of their technique, they require the transcription text to be specified in advance.  To address this need, we use the transcription from Deepspeech ASR with adversarial attacks generated by our technique as Carlini et al.'s  target. We then compare our technique with Carlini et al.~with respect to time taken to generate adversarial attacks,  \texttt{Similarity} to original audio, \texttt{Success Rate} on other ASRs, Google and Sphinx, and \texttt{Detection score}. Since the transcription text in both techniques are the same, it is not useful to compare \texttt{WER}.

We compare our technique against Abdullah et al. using \texttt{WER, Similarity, Success Rate, Detection Score, Time} over different ASRs and both the Commonvoice and Librispeech dataset. 

\paragraph{Experiment settings} We use Google Colab Pro with two NVIDIA Tesla T4 GPUs(16GB RAM, 2560 cores)  to run our experiments. 
We use the following audio parameters in our experiments:
Sampling rate of $16000HZ$, frame length of $2048$ and frame shift of $512$.

\section{Results and Analysis}
\label{sec:results}
We present and discuss the results from our experiments in the context of the research questions presented earlier. 
It is worth noting that \texttt{WER} and \texttt{Similarity} are 
measured for each attack, while \texttt{Success rate} and \texttt{Detection score} are measured across an entire dataset. Techniques should try to maximise \texttt{WER, Similarity} and \texttt{Success rate} while minimising \texttt{Detection score} by Waveguard. 
\begin{table*}[htbp]
\centering
\setlength{\tabcolsep}{3mm}{
    \begin{tabular}{cc}

         Librispeech                                   & Commonvoice \\
         \includegraphics[scale=0.45]{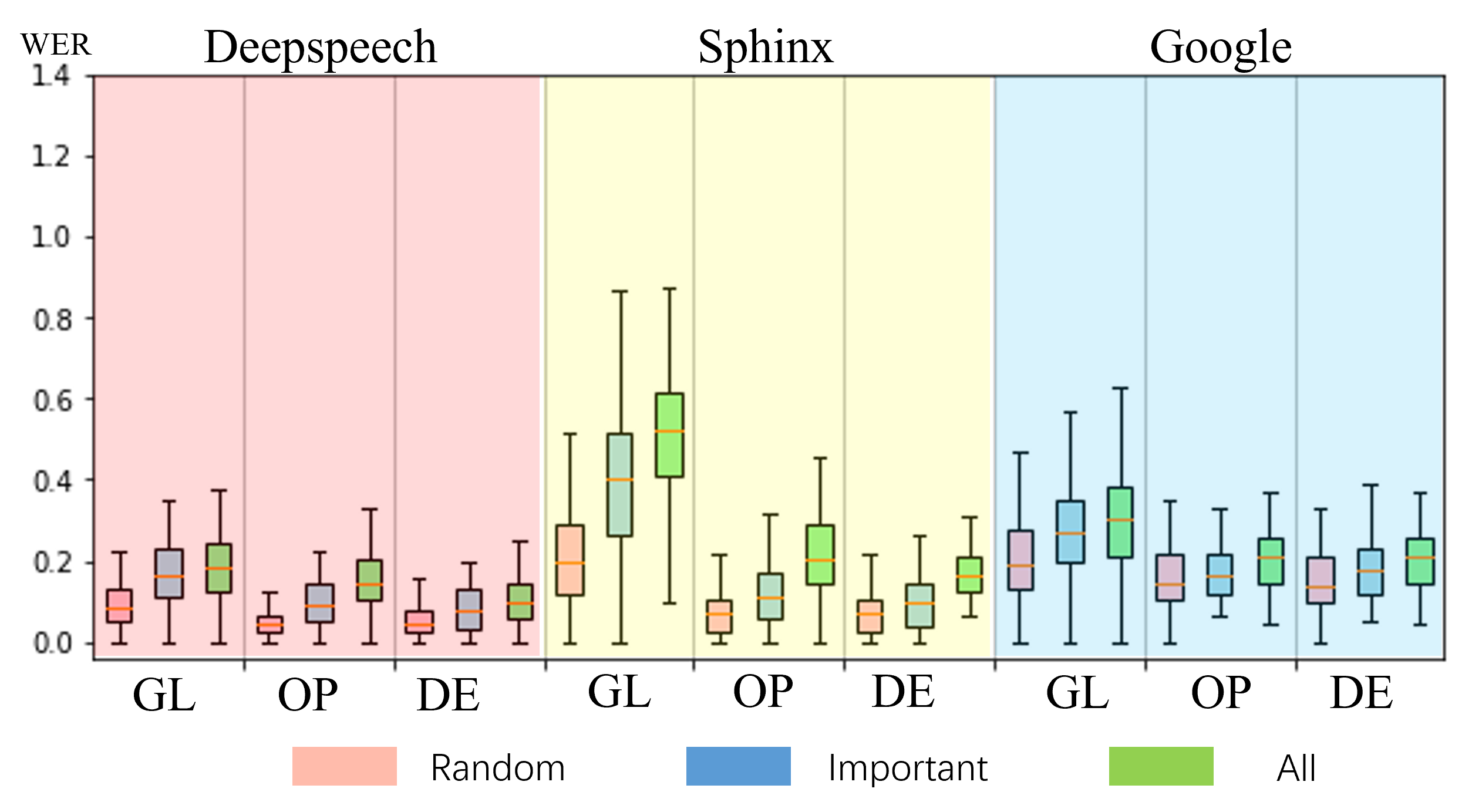} & \includegraphics[scale=0.45]{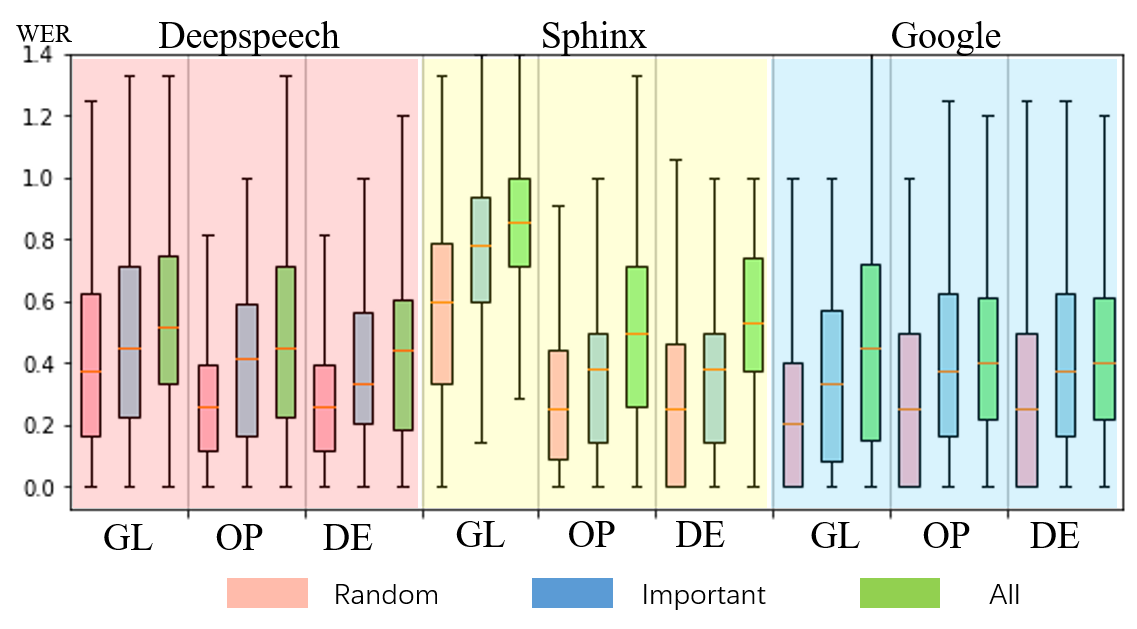}
 \\
    
    \end{tabular}}
\caption{Box plots of the \texttt{WER} of the adversarial attacks generated with two different datasets.}
\label{table: wer}
\end{table*}

\begin{figure}[htbp]
\includegraphics[scale=0.60]{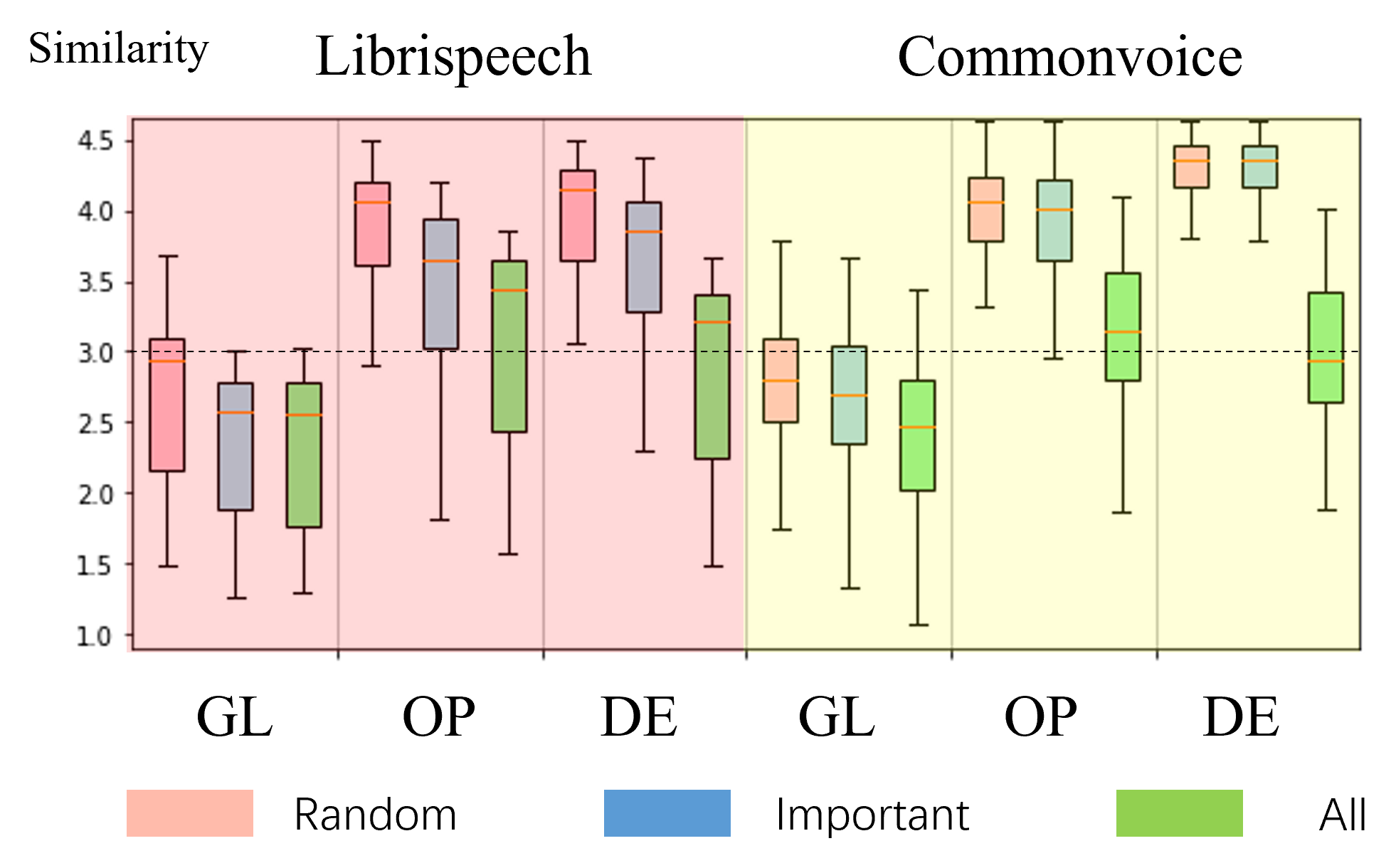}
\caption{Box plots of the \texttt{Similarity} of the adversarial attacks generated with all datasets.}
\label{fig: Similarity}
\end{figure}


\begin{figure}[htbp]
\includegraphics[scale=0.60]{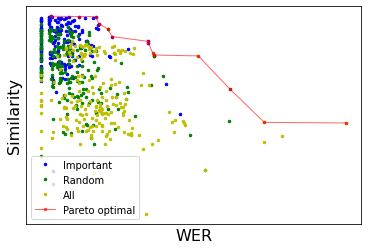}
\caption{Pareto front over adversarial attacks generated by \texttt{Random}, \texttt{Important} and \texttt{All} frame selection techniques on Commonvoice dataset and Deepspeech ASR using \texttt{OP}.}
\label{fig: optimal_frames}
\end{figure}

\begin{figure}[htbp]
\includegraphics[scale=0.60]{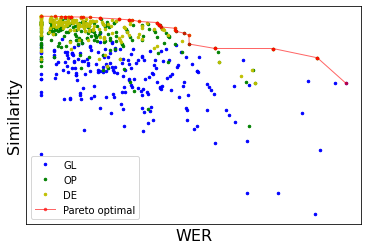}
\caption{Pareto front over adversarial attacks generated by \texttt{GL}, \texttt{OP} and \texttt{DE} on Commonvoice dataset and Deepspeech ASR using \texttt{Important} frames.}
\label{fig: optimal}
\end{figure}

\begin{table*}[htbp]
\centering
\setlength{\tabcolsep}{4.5mm}{
\begin{tabular}{|c|
>{\columncolor[HTML]{FFCCC9}}c 
>{\columncolor[HTML]{FFCCC9}}c 
>{\columncolor[HTML]{FFCCC9}}c |
>{\columncolor[HTML]{FFFC9E}}c 
>{\columncolor[HTML]{FFFC9E}}c 
>{\columncolor[HTML]{FFFC9E}}c |}
\hline
\multicolumn{1}{|l|}{}                   & \multicolumn{3}{c|}{\cellcolor[HTML]{FFCCC9}Librispeech}                                                                                                                              & \multicolumn{3}{c|}{\cellcolor[HTML]{FFFC9E}Commonvoice}                                                         \\ \cline{2-7} 
\multicolumn{1}{|l|}{\multirow{2}{*}} & \multicolumn{1}{c|}{\cellcolor[HTML]{FFCCC9}{\color[HTML]{000000} GL}}   & \multicolumn{1}{c|}{\cellcolor[HTML]{FFCCC9}{\color[HTML]{000000} OP}}     & {\color[HTML]{000000} DE}     & \multicolumn{1}{c|}{\cellcolor[HTML]{FFFC9E}GL}   & \multicolumn{1}{c|}{\cellcolor[HTML]{FFFC9E}OP}     & DE     \\ \hline
{\color[HTML]{000000} Deepspeech}        & \multicolumn{1}{c|}{\cellcolor[HTML]{FFCCC9}{\color[HTML]{000000} 96\%}} & \multicolumn{1}{c|}{\cellcolor[HTML]{FFCCC9}{\color[HTML]{000000} 95\%}}   & {\color[HTML]{000000} 91\%}   & \multicolumn{1}{c|}{\cellcolor[HTML]{FFFC9E}95\%} & \multicolumn{1}{c|}{\cellcolor[HTML]{FFFC9E}90\%}   & 90\%   \\ \hline
{\color[HTML]{000000} Sphinx}            & \multicolumn{1}{c|}{\cellcolor[HTML]{FFCCC9}{\color[HTML]{000000} 99\%}} & \multicolumn{1}{c|}{\cellcolor[HTML]{FFCCC9}{\color[HTML]{000000} 96.5\%}} & {\color[HTML]{000000} 94\%}   & \multicolumn{1}{c|}{\cellcolor[HTML]{FFFC9E}98\%} & \multicolumn{1}{c|}{\cellcolor[HTML]{FFFC9E}89\%}   & 90\%   \\ \hline
{\color[HTML]{000000} Google}            & \multicolumn{1}{c|}{\cellcolor[HTML]{FFCCC9}{\color[HTML]{000000} 99\%}} & \multicolumn{1}{c|}{\cellcolor[HTML]{FFCCC9}{\color[HTML]{000000} 97.5\%}} & {\color[HTML]{000000} 95.5\%} & \multicolumn{1}{c|}{\cellcolor[HTML]{FFFC9E}85\%} & \multicolumn{1}{c|}{\cellcolor[HTML]{FFFC9E}80\%}   & 80\%   \\ \hline
Average                                  & \multicolumn{1}{c|}{\cellcolor[HTML]{FFCCC9}98\%}                        & \multicolumn{1}{c|}{\cellcolor[HTML]{FFCCC9}96.3\%}                        & 93.5\%                        & \multicolumn{1}{c|}{\cellcolor[HTML]{FFFC9E}92\%} & \multicolumn{1}{c|}{\cellcolor[HTML]{FFFC9E}86.3\%} & 86.6\% \\ \hline
\end{tabular}}
\caption{The  \texttt{Success Rate}s of the adversarial attacks with \texttt{GL,OP,DE} attack generation methods across the three ASRs and two datasets.\texttt{All} frames is used as the frame selection method.}
\label{table: success_rate}
\end{table*}

\begin{table*}[htbp]
\centering
\setlength{\tabcolsep}{1mm}{
\begin{tabular}{|c|c|c|ccl|ccl|c|}
\hline
{Technique} & {Time} & {Similarity} & \multicolumn{3}{c|}{Success rate}                                      & \multicolumn{3}{c|}{WER}                                               & {Detection score} \\ \cline{4-9}
                           &                       &                             & \multicolumn{1}{c|}{Deepspeech} & \multicolumn{1}{c|}{Sphinx} & Google & \multicolumn{1}{c|}{Deepspeech} & \multicolumn{1}{c|}{Sphinx} & Google &                            \\ \hline
\textbf{Carlini~\cite{carlini2018}}          & 780 seconds            & 3.63                        & \multicolumn{1}{c|}{N/A}       & \multicolumn{1}{c|}{77\%}   & 33\%   & \multicolumn{1}{c|}{N/A }       & \multicolumn{1}{c|}{N/A}   & {N/A}    & 0.67                       \\ \hline
\textbf{Abdullah~\cite{abdullah2019hear}}          & 18 seconds            & 3.12                        & \multicolumn{1}{c|}{80\%}       & \multicolumn{1}{c|}{77\%}   & 54\%   & \multicolumn{1}{c|}{0.39}       & \multicolumn{1}{c|}{0.44}   & 0.14   & 0.65                       \\ \hline
\textbf{OP+Important}      & 155 seconds           & 3.93                        & \multicolumn{1}{c|}{86\%}       & \multicolumn{1}{c|}{78\%}   & 75\%   & \multicolumn{1}{c|}{0.41}       & \multicolumn{1}{c|}{0.41}   & 0.39   & {\color[HTML]{FE0000}0.52}                       \\ \hline
\textbf{OP+All}            & 3.5 seconds           & 3.22                        & \multicolumn{1}{c|}{{\color[HTML]{FE0000}90\%}}       & \multicolumn{1}{c|}{89\%}   & {\color[HTML]{FE0000}80\%}   & \multicolumn{1}{c|}{{\color[HTML]{FE0000}0.44}}       & \multicolumn{1}{c|}{0.47}   & {\color[HTML]{FE0000}0.40}   & 0.53                       \\ \hline
\textbf{DE+Important}      & 154 seconds           & {\color[HTML]{FE0000}4.29}                        & \multicolumn{1}{c|}{84\%}       & \multicolumn{1}{c|}{77\%}   & 74\%   & \multicolumn{1}{c|}{0.39}       & \multicolumn{1}{c|}{0.40}   & 0.36   & 0.55                       \\ \hline
\textbf{DE+All}            & {\color[HTML]{FE0000}2.5 seconds}           & 3.13                        & \multicolumn{1}{c|}{{\color[HTML]{FE0000}90\%}}       & \multicolumn{1}{c|}{{\color[HTML]{FE0000}90\%}}   & {\color[HTML]{FE0000}80\%}   & \multicolumn{1}{c|}{{\color[HTML]{FE0000}0.44}}       & \multicolumn{1}{c|}{{\color[HTML]{FE0000}0.50}}   & 0.38   & 0.56                       \\ \hline
\end{tabular}}
\caption{Comparison of \texttt{OP+All, OP+Important, DE+All, DE+Important} with Abdullah et al.~\cite{abdullah2019hear} and Carlini et al.~\cite{carlini2018} with respect to generation time for per adversarial attack, \texttt{Similarity} to original audio examples,\texttt{WER}, \texttt{Success Rate} and \texttt{Detection score} against defense system~\cite{waveguard} in attacking all three ASRs}
\label{comparation_abdul}
\end{table*}
\subsection{RQ1: Comparison of Frame Selection Techniques}
The best performing frame selection technique is one that achieves high \texttt{WER} and high \texttt{Similarity} across audio examples. However, these two metrics are often conflicting. We discuss and compare \texttt{WER} and \texttt{Similarity}  achieved by the three frame selection techniques in our approach below.  Figures in Table~\ref{table: wer} shows the \texttt{WER} achieved by different frame section techniques for the Librispeech and Commonvoice datasets across different ASRs and attack generation techniques while Figure~\ref{fig: Similarity} shows the \texttt{Similarity} achieved. 
\paragraph{\texttt{All} frames} We find in Table~\ref{table: wer} and Figure~\ref{fig: Similarity}, that the \texttt{All} frame selection achieves the highest \texttt{WER} and lowest \texttt{Similarity} compared to \texttt{Important} and \texttt{Random} across ASRs, input datasets and attack generation methods.  
This is in line with our expectations as the other two frame selection techniques select a small part of the audio to introduce noise into achieving lower \texttt{WER} but higher \texttt{Similarity} to original audio.

\paragraph{\texttt{Important} versus \texttt{Random}:} For most combinations of ASR, dataset and attack generation, we find \texttt{Random} frame selection produces the lowest \texttt{WER} and the highest \texttt{Similarity}, while \texttt{Important} frame selection results in a \texttt{WER} and \texttt{Similarity} between \texttt{Random} and \texttt{All}. 

\paragraph{Statistical Analysis.} We confirmed the statistical significance (at 5\% significance level) of the difference in means between the frame selection techniques using one-way Anova and did a post-hoc Tukey's Honest Significant Difference (HSD) test to reveal which differences between pairs of means are significant. Supplementary material Sections 1.1.1 and 1.1.2 list the P-values for pairwise comparisons of \texttt{WERs} and \texttt{Similarities} between frame selection techniques. 
For the \texttt{WER} metric, we find the \texttt{All} frames selection technology is significantly better than \texttt{Important} and \texttt{Random} on majority of ASR, dataset, attack technique combinations. In contrast, for \texttt{Similarity} measure, \texttt{Random} and \texttt{Important} frame selections significantly outperformed \texttt{All}. 
\paragraph{Pareto front} 
Owing to the conflicting nature of the \texttt{WER} and \texttt{Similarity} metrics, all three frame selection techniques achieve a trade-off between them. We use the Pareto front with these two metrics, shown in Figure~\ref{fig: optimal_frames} for one of the datasets and ASRs, to determine the number of non-dominated attack examples (that fall on the Pareto front) from each frame selection. 
We find \texttt{Important} frame selection has the most number of non-dominated attacks (25 examples); \texttt{Random} was second with 15 examples, while \texttt{All} frames only had 1 non-dominated attack example. This trend is observed across all ASRs, attack technologies and datasets (see results in Supplementary material Section 1.1.3).  Based on the number of non-dominated examples, we believe that \texttt{Important} frames is effective at achieving a trade-off between \texttt{WER} and \texttt{Similarity}.

\paragraph{Summary} In terms of \texttt{WER}, we find \texttt{All} frames performs best. However, \texttt{Important} and  \texttt{Random} frames perform better in terms of \texttt{Similarity}. 
We find \texttt{Important} is the best at optimising trade-off between the two metrics, achieving reasonable performance in both \texttt{WER} and \texttt{Similarity}. 


\subsection{RQ2: Comparison of Attack Generation Techniques} 
We present \texttt{WER} achieved by \texttt{GL, OP, DE} using different ASRs and datasets in Table~\ref{table: wer}, while we show \texttt{Similarity} achieved 
in Figure~\ref{fig: Similarity}. 
Best performing attack generation technique is one that results in a high \texttt{WER} and high \texttt{Similarity} to original audio.


\paragraph{WER Performance} \texttt{GL} attack generation performs better than both \texttt{OP} and \texttt{DE} in terms of WER achieved. We confirm the differences are significant using One-way Anova and Tukey's HSD test (see P-values in Section 1.2.1 of the Supplementary material). Between \texttt{OP} and \texttt{DE} attacks, \texttt{OP} outperforms \texttt{DE} with DeepSpeech and Sphinx ASRs over the Librispeech dataset.
There is no significant difference between the two techniques over the other dataset and ASRs.

\paragraph{Similarity Performance} Both \texttt{OP} and \texttt{DE} significantly outperform \texttt{GL} in terms of \texttt{Similarity}, confirmed with pairwise comparison using one-way Anova followed by Tukey's HSD test (P-value tables in Supplementary material Section 1.2.2). The median \texttt{Similarity} or PESQ score for \texttt{GL} tends to be below the value of $3.0$ (shown by the dashed line), irrespective of frame selection used. According to Beuran et al.~\cite{PESQover3}, the standard for good quality audio is a PESQ score of greater than $3$ and \texttt{GL} technique does not meet this standard in our experiments. We believe this is because \texttt{GL} uses estimated, rather than actual,  phase information which causes distortion that reduces the PESQ score.  

Between OP and DE, there is no significant difference in their \texttt{Similarity} performance. The benefit with using \texttt{DE} lies in faster generation of an adversarial attack. The average time to generate a single adversarial attack using \texttt{DE} is $2.5 seconds$, a second faster than the \texttt{OP} technique ($3.5 seconds$ on average) as \texttt{OP} relies on calculating the masking threshold for every input example. 

\paragraph{Pareto Front} 
As with RQ1, we draw the Pareto front using \texttt{WER} and \texttt{Similarity}, shown in Figure~\ref{fig: optimal}. We find \texttt{DE} technique has the most number of non-dominated attacks (28 examples); \texttt{OP} is second with 10 examples, while \texttt{GL} only has 1 non-dominated attack example. This trend is observed across all ASRs, frame selections and datasets (Results available in Section 1.1.3 of the Supplementary material).

\paragraph{Summary}
Based on the number of non-dominated examples, we believe that \texttt{DE} is a suitable choice for optimising both \texttt{WER} and \texttt{Similarity}. Additionally, 
\texttt{DE} is the fastest attack generation technique. Taking both these aspects into account, we believe \texttt{DE} would be the best choice for attack generation. 
\subsection{RQ3: Portability across ASRs}
We evaluate portability of the adversarial attacks generated by \texttt{OP,GL,DE} across the three ASRs using the \texttt{Success Rate} metric, described in Section~\ref{sec:metrics}.  
Table~\ref{table: success_rate} presents \texttt{Success Rates} achieved with the Librispeech and Commonvoice datasets.

We find \texttt{GL} achieves the best success rates over all ASRs, with both the Librispeech dataset (average of $98\%$) and the Commonvoice dataset (average of $92\%$). \texttt{OP} comes next, performing better than \texttt{DE} on the Librispeech dataset ($96\%$ versus $93.5\%$, respectively). \texttt{OP} and \texttt{DE} have similar performance over the Commonvoice dataset (average of $86\%$). 

\paragraph{Summary} All three attack generation techniques have high success rates across the three ASRs producing portable adversarial attacks. \texttt{GL} outperforms \texttt{OP} and \texttt{DE} in portability but the magnitude of difference is small (on average $2\%$ to $5\%$). \texttt{OP} and \texttt{DE} have comparable performance on the ASRs, especially with the Commonvoice dataset.

\subsection{RQ4: Comparison to Existing Techniques}
\label{sec:RQ4}
As mentioned in Section~\ref{sec:RQ}, we compare performance of our approach against a whitebox targeted technique proposed by Carlini et al.~\cite{carlini2018} and a blackbox untargeted technique proposed by Abdullah et al.~\cite{abdullah2019hear} using the metrics -- \texttt{WER, Similarity, Success rate, Time, Detection score}. 

\subsubsection{Comparison with Carlini et al} We fix the ASR to Deepspeech and input dataset to Commonvoice to match the experiments in Carlini et al.~\cite{carlini2018}. For comparison, we use the best performing techniques in our approach (for \texttt{Similarity} and \texttt{WER}) -- \texttt{OP} and \texttt{DE} for attack generation with \texttt{Important} and \texttt{All} frame selections. We show results in Table~\ref{comparation_abdul}. We do not compare \texttt{WER} as the target text for Carlini et al.~\cite{carlini2018} is the transcription text from our adversarial attacks, so there will be no difference. 

\paragraph{\texttt{Time} and \texttt{Similarity}} We find time taken to generate attack examples is faster with our approaches, \texttt{OP} and \texttt{DE}, compared to Carlini et al. with a maximum speedup of $312 \times$ achieved with \texttt{DE+All}. 
We also achieve higher \texttt{Similarity} scores when using \texttt{Important} frames -- $4.3$ (\texttt{DE+Important}) and $3.9$ (\texttt{OP+Important}), compared to $3.6$ by Carlini et al. We confirm the statistical significance (at 5\% significance level) of the observed differences in \texttt{Similarity} using one-way Anova and Tukey's Honest Significant Difference (HSD) test.
We find our techniques are a clear winner in terms of time taken, and outperform Carlini at al. in \texttt{Similarity} when using \texttt{Important} frames but not \texttt{All} frames. \texttt{Similarity} performance difference between \texttt{Important} and \texttt{All} was discussed in RQ1. 

\paragraph{Success Rate} To evaluate portability of adversarial attacks, we transcribe the adversarial attacks using Google and Sphinx (since DeepSpeech is used by Carlini et al.). We find when used with Google ASR, adversarial attacks generated by Carlini et al.~have a much lower \texttt{Success Rate} than our techniques (33\%  versus 74\% to 80\%), respectively. For Sphinx, the difference in \texttt{Success Rate} is smaller but the trend remains (77\% Carlini versus 77\% to 90\% for ours). 
The lower \texttt{Success Rate} observed with Carlini et al.~is because their technique specifically targets  the neural network inside Deepspeech, and may not be as effective when used on other ASRs with different NNs. This is a drawback also encountered with other whitebox attacks. However,  since our method is blackbox, we find it is easier to port our adversarial attacks to different ASRs. 

\paragraph{Detection score} 
Attack examples generated by Carlini et al. are more easily detected by Waveguard, with a higher \texttt{Detection score} score of $0.67$, compared to techniques in our approach, whose \texttt{Detection score} range from $0.52$ to $0.56$. We believe this is because Carlini et al use noise in their attack generation which is detected more easily by Waveguard. 
We find the four techniques in our approach perform better than Carlini et al at evading the Waveguard defense. 

Across all four evaluation metrics, we find one of the four techniques from our approach is the winner (highlighted in red in Table~\ref{comparation_abdul}), outperforming Carlini et al. Among them, \texttt{OP+Important} and \texttt{DE+Important} is superior to Carlini et al.~\cite{carlini2018} across all metrics. \texttt{OP+All} and \texttt{DE+All} show significant gains in generation time and  \texttt{Success Rate} but at the cost of \texttt{Similarity} which is slightly lower than Carlini et al. 

\subsubsection{Comparison with Abdullah et al} Like our approach, Abdullah et al.~\cite{abdullah2019hear}  use a blackbox, untargeted attack generation technique that is meant to be fast and portable on different ASRs. 
Unlike the comparison with Carlini et al., we can include \texttt{WER} as a performance metric (in addition to the other 4 metrics) and Deepspeech ASR in our comparison. We discuss performance for each of the metrics below using the Commonvoice dataset\footnote{Results for Librispeech dataset follow a similar trend and can be viewed in Supplementary material Section 1.3.2.}.

\paragraph{\texttt{Time} and \texttt{Similarity}} We find our approach, \texttt{OP} and \texttt{DE} with \texttt{All} frames, is much faster in generating attacks than Abdullah et al. ($5 \times$ and $7 \times$ faster, respectively). In contrast, Abdullah et al. is $8$ times faster than \texttt{OP} and \texttt{DE} when they use \texttt{Important} frames, where much of the time with our approach is spent in frame selection.   
For the \texttt{Similarity} metric, our approach outperforms Abdullah et al. with all 4 techniques (at 5\% significance level, P-value tables in the Supplementary material.) As noted in RQ1, \texttt{Important} frame selection achieves better \texttt{Similarity} scores than \texttt{All} frames. 

\paragraph{\texttt{Success rate}, \texttt{WER} and \texttt{Detection score}} Attack examples generated with \texttt{OP} and \texttt{DE} have a higher \texttt{Success rate} than Abdullah et al. across all ASRs. Selecting \texttt{All} frames with our attack techniques achieves the best \texttt{Success rate}. We see a similar trend with \texttt{WER}, where \texttt{OP} and \texttt{DE} outperform Abdullah et al. (at 5\% statistical significance). Finally, \texttt{OP} and \texttt{DE} surpass Abdullah et al. with respect to getting past Waveguard's defense system by achieving lower detection scores of $0.52 - 0.56$ versus $0.65$ for Abdullah et al.. 

In summary, we find our attack techniques, \texttt{OP} and \texttt{DE}, surpass Abdullah et al. for each of the five evaluation metrics (best performing is highlighted in red in Table~\ref{comparation_abdul}). Choice of frame selection within \texttt{OP} and \texttt{DE} impacts attack generation \texttt{Time} and \texttt{Similarity} while the relative performance on the remaining metrics is largely unaffected. 


\section{Related Work}
\begin{table}[htbp]
\centering
\begin{tabular}{|p{3cm}|p{4.5cm}|}
\hline
Attack Type             & Existing work \\\hline
Whitebox-Targeted   &  Vaidya et al.~\cite{Vaidya_cocaine}, Carlini et al,~\cite{carlini2016,carlini2018}, Qin et al.~\cite{qin2019},Yuan et al.~\cite{yuancommandersong},Yakura et al.~\cite{Yakurarobust}, Schönherr et al.~\cite{schonherr2018,schonherr2020imperio}, Szurley et al.~\cite{szurley2019perceptual}     \\\hline
Blackbox-Targeted   &  Zhang et al.~\cite{Zhang2017DolphinAttack}, Alzantot et al.~\cite{didyouhear}, Taori et al.~\cite{taori2019targeted}     \\\hline
Blackbox-Untargeted &    Abdullah et al.~\cite{abdullah2019hear}   \\\hline
\end{tabular}
\caption{Existing work on adversarial ASR attack generations.}
\vspace{-10pt}
\label{existing work}
\end{table}

As mentioned in Section~\ref{sec:intro}, existing adversarial attack generation on ASR models can be classified along two dimensions: 1. Targeted for a given transcription or untargeted, and 2.  Whitebox, with knowledge of the internal ASR structure or Blackbox. Table~\ref{existing work} lists the existing techniques using these two dimensions and they are discussed in more detail in the rest of this Section.   

\subsection{Targeted Attacks}
\label{sec:targeted}

Vaidya et al.~\cite{Vaidya_cocaine} pioneered the first whitebox targeted method for attacking ASR in 2015. Given the transcription to target, they gradually approach the target by continuously fine-tuning the parameters of the extracted MFCC features. Once the goal is reached, they use the obtained adversarial MFCC features to reconstruct the speech waveform. On the basis of Vaidya's work and in an effort to improve the efficiency of their approach, Carlini et al.\cite{carlini2016} proposed Hidden Voice Command in 2016, adding noise that is often encountered in real life. However, neither of these two types of attacks can conceal the existence of noise, and such adversarial attacks can be easily detected as noise rather than effective commands. 

Yuan et al.~\cite{yuancommandersong} proposed a method for embedding commands into songs so that when these songs are played, the commands will be translated by an ASR. Additionally, they improve the realistic nature of adversarial attacks by introducing noise generated by hardware devices. This approach, however, is restricted to songs as the carrier of commands, and is, therefore, limited in application scenarios. 

Carlini et al.~\cite{carlini2018} in 2018 used a whitebox approach that applies gradient descent to modify the original audio so that the difference between the transcription and the target text is smaller. Their experimental results show their attack  \texttt{Success Rates} reached 100\% on Deepspeech ASR. 
However, their approach faces the following drawbacks: First, it can take up to several hours to generate attacks; second, the gradient descent method requires the attacker to have a good understanding of all the internal parameters and structures of the attacked system before it can be used; and finally the adversarial attacks generated will be invalid over other ASRs. 

Yakura et al.~\cite{Yakurarobust} proposed some improvements to~\cite{carlini2018} to maintain attack performance under over-the-air conditions (mixed with sound of the surrounding environment). They generate adversarial attacks accounting for noise caused by echo and recording in real life, so as to obtain more robust adversarial attacks. However,  other shortcomings in Carlini et al.\cite{carlini2018} (such as long generation time and weak transferability) have not been addressed. 

In 2018, Schönherr et al.~\cite{schonherr2018} developed a whitebox approach that applies the knowledge of masking threshold to generate adversarial attacks. They proposed to limit the generated noise below the masking threshold of the original audio to ensure that the obtained perturbation is not audible to the human ear. In more recent work~\cite{schonherr2020imperio}, they introduced room impulse response (RIR) simulator to improve the robustness of examples that produces different types of noise for different environment configurations. 


Inspired by Schönherr et al., Qin and Carlini et al.~\cite{qin2019} developed a whitebox method and optimized perturbations to make it lower than the masking threshold of the original audio. This method achieved a 100\% attack  \texttt{Success Rate} on the Lingvo system. 
Like other whitebox targeted approaches, their work lacks portability to other ASRs and is time consuming for attack generation. 

Around the same time, Szurley et al.~\cite{szurley2019perceptual} proposed a whitebox method similar to Schönherr et al.~\cite{schonherr2018, schonherr2020imperio}  
and Carlini et al.~\cite{qin2019, carlini2018} that constructed an optimization based on masking threshold and combined it with room reverberation. Their method reached a 100\%  \texttt{Success Rate} on Deepspeech but still suffers from limitations of lack of portability and time consuming attack generation.
\paragraph{Blackbox-targeted approaches}
Few Blackbox Targeted adversarial attack generation techniques exist in the literature~\cite{Zhang2017DolphinAttack,didyouhear,taori2019targeted}. 
Zhang et al.~\cite{Zhang2017DolphinAttack} in 2017 
modulated the voice on the ultrasonic carrier to 
insert preset commands(like "Open the window") into the original audio. 
However, this method is not easy to reproduce as it uses hardware characteristics of the microphone to complete the attack. 
Alzantot et al.~\cite{didyouhear} proposed a iterative optimization method that adds a small amount of noise iteratively to a benign example until the ASR outputs a target label. Taori et al.~\cite{taori2019targeted} used a genetic algorithm to achieve iterative optimization, mutating benign examples until the ASR output matches a target label. These approaches for blackbox targeted attacks suffer from the following two weaknesses: First, they require thousands of queries to ASRs to generate one adversarial attack, which is unrealistic. Secondly, these attacks are only applicable to ASRs that aim to classify audios, not translate audios.

\subsection{Untargeted Attacks}
\label{sec:untargeted}
The only known untargeted blackbox adversarial ASR attack generation approach is that proposed by 
Abdullah et al.~\cite{abdullah2019hear} in 2019. They construct an adversarial attack by decomposing and reconstructing the original audio. Specifically, they decompose the original audio into components called eigenvectors via Singular Spectrum Analysis (SSA). These eigenvectors represent the various trends and noises that make up the audio. They believe that eigenvectors with smaller eigenvalues convey limited information. They choose a threshold to classify eigenvalues as small and subsequently eliminate small eigenvectors. They then reconstruct an audio from the remaining components as the adversarial attack. 
We compare performance of our techniques against their approach in Section~\ref{sec:RQ4}.

\section{Conclusion}
We proposed a blackbox untargeted adversarial attack generation technique for ASRs using frequency masking to make the adversarial audio sound similar to the original while producing a change in the transcription. Our approach  provides three attack generation options  -- \texttt{GL, OP} and \texttt{DE}.  We  also  provide  the  option  of  selectively  introducing perturbations to a small fraction of audio frames using three frame selection options –- \texttt{Random, Important} and \texttt{All}. Evaluation of our techniques over three ASRs and two audio datasets showed that our techniques can be effective at achieving high \texttt{WERs} (average of $44\%$ with \texttt{OP+All}) while also achieving high \texttt{Similarity} (average of $3.93$ with \texttt{OP+Important}). The choice in attack generation and frame selection helps achieve a good balance between these two metrics, with \texttt{DE} attack generation and \texttt{Important} frames achieving the best trade-off. We also confirmed that our techniques were portable across ASRs and superior to existing whitebox targeted technique~\cite{carlini2018} and blackbox untargeted technique~\cite{abdullah2019hear} in terms of \texttt{WER, Similarity, Success Rate, Time} and \texttt{Detection score}. 

\bibliographystyle{ACM-Reference-Format}
\bibliography{sample-base}

\end{document}